\documentclass{vgtc}                          

\ifpdf
  \pdfoutput=1\relax                   
  \usepackage{graphicx}                
  \DeclareGraphicsExtensions{.pdf,.png,.jpg,.jpeg} 
\else
  \usepackage{graphicx}                
  \DeclareGraphicsExtensions{.eps}     
\fi%

\graphicspath{{figures/}{pictures/}{images/}{./}} 

\usepackage{microtype}                 
\PassOptionsToPackage{warn}{textcomp}  
\usepackage{textcomp}                  
\usepackage{mathptmx}                  
\usepackage{times}                     
\usepackage{cite}                      
\usepackage{tabu}                      
\usepackage{booktabs}                  

\onlineid{0}

\vgtccategory{Research}

\vgtcinsertpkg

\title{The Mood of the Sunlight: A Public Artwork based on Visualization of Sunlight Data}

\author{Yifan Wang\thanks{e-mail: 119010317@link.cuhk.edu.cn}\\ %
        \parbox{1.8in}{\scriptsize \centering School of Data Science, The Chinese University of Hong Kong, Shenzhen \\ Shenzhen Institute of Artificial Intelligence and Robotics for Society} %
\and Nan Li\thanks{e-mail: linan@cuhk.edu.cn}\\ %
    \parbox{1.8in}{\scriptsize \centering Shenzhen Institute of Artificial Intelligence and Robotics for Society \\ Institute of Robotics and Intelligent Manufacturing, The Chinese University of Hong Kong, Shenzhen}  %
\and Suxuan Jiang\thanks{e-mail: suxuan\_J@163.com}\\ %
     \parbox{1.8in}{\scriptsize \centering School of Architecture \& Urban Planning, Shenzhen University}\\
\and Jinlong Xu\thanks{e-mail: 316731786@qq.com}\\ %
     \parbox{1.8in}{\scriptsize \centering School of Architecture \& Urban Planning, Shenzhen University}\\
\and Qi Wang\thanks{e-mail: 119010302@link.cuhk.edu.cn}\\ %
    \parbox{1.8in}{\scriptsize \centering School of Data Science, The Chinese University of Hong Kong, Shenzhen \\ Shenzhen Institute of Artificial Intelligence and Robotics for Society}\\
\and Shaomin Shen\thanks{e-mail: shaomin\_shen@163.com}\\ %
     \parbox{1.8in}{\scriptsize \centering School of Architecture \& Urban Planning, Shenzhen University}\\
\and Ning Ding\thanks{e-mail: linan@cuhk.edu.cn}\\ %
    \parbox{1.8in}{\scriptsize \centering Shenzhen Institute of Artificial Intelligence and Robotics for Society \\ Institute of Robotics and Intelligent Manufacturing, The Chinese University of Hong Kong, Shenzhen}
     }

\abstract{The application of data visualization in public art attracts increasing attention. In this paper, we present the design and implementation of a visualization method for sunlight data collected over a long period of time with an industrial camera. The proposed method makes use of the saturation and value information of collected sunlight image data in Hue Saturation Value (HSV) color model to show the variation of the “mood” of the sunlight. Specifically, we create visual patterns with a rotating planet gear, which has an intuitively consistent geometric meaning with HSV color model and the planetary motion. Due to the variation of the sunlight data over time, the generated visual pattern presents a periodic variation that corresponds to the changing “mood” of the sunlight. Furthermore, we also use the sunlight data to generate music as another form of data representation. Two public artworks have been created with the above visualization and auralization methods and displayed on an exhibition held at China Resources Tower, Shenzhen, China. This work is a typical practice of creating public installations with data visualization technology, giving a glimpse into the many ways science and art intersect.      
The two public artworks generated in this work can been seen at https://github.com/crowang1A/The-Mood-of-Sunlight.%
} 

\CCScatlist{
\CCScatTwelve{Visualization and data mining}{Multimedia(Image/Video/Music)}{}{};
\CCScatTwelve{General topics and techniques}{Animation)}{}{};
\CCScatTwelve{General topics and techniques}{Presentation, Production, and Dissemination}{}{}
  }

\begin{document}


\firstsection{Introduction}

\maketitle

\textit{The farther, the art becomes more scientific, and science more artistic: having parted at the base, they will meet someday at the top. - Gustave Flaubert.}

Art and science are both innovative, but are often regarded as distinct. (When we talk about science, we mean an extended scope that includes both the narrowly defined science and technology.) Conventionally, they engage two separate groups of people: scientists and engineers undertake scientific research and development works, while artists take part in artistic creation. In recent years, the integration of art and science has become increasingly popular in the world. Informationization and digitalization are progressing rapidly in the field of art, which can be intuitively felt through the ubiquitous digital media art \cite{Zhang:2021:AFI}\cite{Qu:2022:DVM} and the recent hotspot AI-Generated Art \cite{Li:2022:AI}\cite{Zou:2021:SNP}. Art shows people's infinite imagination; The development of science provides various possibilities for the realization of art. It is of great significance to explore the integration of science and art, conforming to the trend of development.

In this paper, we present our practice of creating a public artwork to explore possible forms of integrating science and public art. We choose the sunlight as the basic element of our artwork, associating its variation with the change of human mood to endow the sun with human motion. We present the variation of the “mood” of sunlight with data visualization technology inspired by planets' movement and geometrical principle, enabling the audience to feel the sunlight from a different dimension and obtain a pleasant sensory experience. Moreover, we propose an auralization method to establish a mapping relationship between the sunlight data with musical notes to generate music so that the audience can listen to the “mood” of sunlight. After artistic processing, i.e., adding background animation and soundtrack, the proposed visualization and auralization methods generated two pieces of public artworks displayed on an exhibition, demonstrating a typical practice that melts science into art.  

\section{RELATED WORKS}

Data visualization is a statistical way to represent information graphically and have been applied to various fields, such as education, medical care, and public decision-making. Since graphs that vary in sizes, shapes, and colors will be used, data visualization not only has statistical significance but also can possess aesthetic value. Nowadays, data visualization has been developed as an independent subject rather than a simple tool, and it has been applied to more and more fields. 

Visualization can be classified into pragmatic visualization and artistic visualization according to the aesthetic criteria \cite{Kosara:2007:VCI}. Pragmatic visualization focuses on daily applications, industry applications and scientific applications, while artistic focuses on visualization applied in design and artworks.

Here are some typical examples of pragmatic visualization. Ceccarini et al. \cite{Ceccarini:2021:EDV} combine data visualization with IoT to create an interactive dashboard that facilitates the management of the campus premises and the timetabling. Bujari et al. \cite{Bujari:2020:MSV} propose a new paradigm to crowd-sourcing the data measurement and visualization tasks to smartphones and other popular smart wearables. O'Handley et al. \cite{OHandley:2022:CPV} develop a data visualization analytic tool for exploring and analyzing students’ progress through a college curriculum.

Some researchers point out that visualization should bridges the gap between design, art, and technical pragmatic information. Previous works have studied the applications of data visualization in art. Some researchers tied color theories to data visualization and develop systems and tools for creating color maps \cite{Brewer:2022:CPM} \cite{Healey:2003:CEC}. Samsel et al. \cite{Samsel:2018:AAE} come up with a model to combine the color mode in ancient painting with data visualization to make it aesthetically attractive. Saito \cite{Saito:2002:FR} creates an artwork entitled “Financial Reminiscences” by visualizing the portfolio texture indicates of financial data. Brunet \cite{Brunet:2014:EDA} proposes a mapping art called “Electrotravelgram” that presents a control panel displaying visualized data collected during the 2-day trip on the sea and river to tell a story of this journey. Samanci et al. \cite{Samanci:2017:FOO} create an interactive art installation that visualizes and auralizes the motion data of sharks and action data of human by controlling the light and sound of three life-size shark skeletons trapped in an ocean made of fiber optic threads.  Liu et al. \cite{Liu:2022:IDV} proposed an audiovisual interaction bio-art installation by visualizing and auralizing Escherichia coli DNA sequence data. Yu \cite{Yu:2016:APV} analyses the physical visualization in artistic creation, proposes to classify physical visualization by whether it is interactive or is programming or not, and analyses how these changes affect to artistic creation. Schroeder et al. \cite{Schroeder:2016:VBS} develop an artist's interface for creating multivariate time-varying data visualizations. 

Our research is a practice to apply data visualization on public art by mapping the color data of sunlight into geometrical motion which dynamically generates curves to reflect the variation of sunlight. By associating the variation of sunlight data with the change of human mood, we endow the sunlight with human emotion. We also auralize the sunlight data to provide another dimension for experience the “mood” of the sunlight. We hope that the resulting artwork will offer a pleasant sensory experience for the audiences and provide a simple but typical reference for the convergence of art and science.

\section{OVERALL DESIGN}

Since ancient times, human beings have always maintained a strong interest in the sun. The sun is regarded as significant theological worship which is assigned to artistic value and derives from plenty of illustrious artworks and characters, such as Apollo, the sun god in ancient Greek, and Xi He, the sun goodness in ancient China. At the same time, the sun is the most important astronomical body in the solar system, it provides the earth with endless solar energy and maintains the stability of the solar system. 

Sometimes, the sunlight is associated with emotion, just as people gain positive emotion when the weather changes from a cold rainy day to a warm sunny day. Therefore, in this research, we collect image data of sunlight to capture its variation over time, and propose a visualization scheme for the collected data. We map the saturation and value of the sunlight data in HSV color model to the geometric dimensions and motion of a planet gear which generates curves when it rotates to present the variation of the data. Furthermore, we propose an auralization scheme by establishing a corresponding relationship between the data and musical notes. By adding appropriate background animation to the video generated from the data visualization and auralization results, two artworks are produced.

\section{DATA COLLECTION AND VISUALIZATION}

\subsection{Data Collection}

\begin{figure}[h]
 \centering
 \includegraphics[width=0.6\columnwidth]{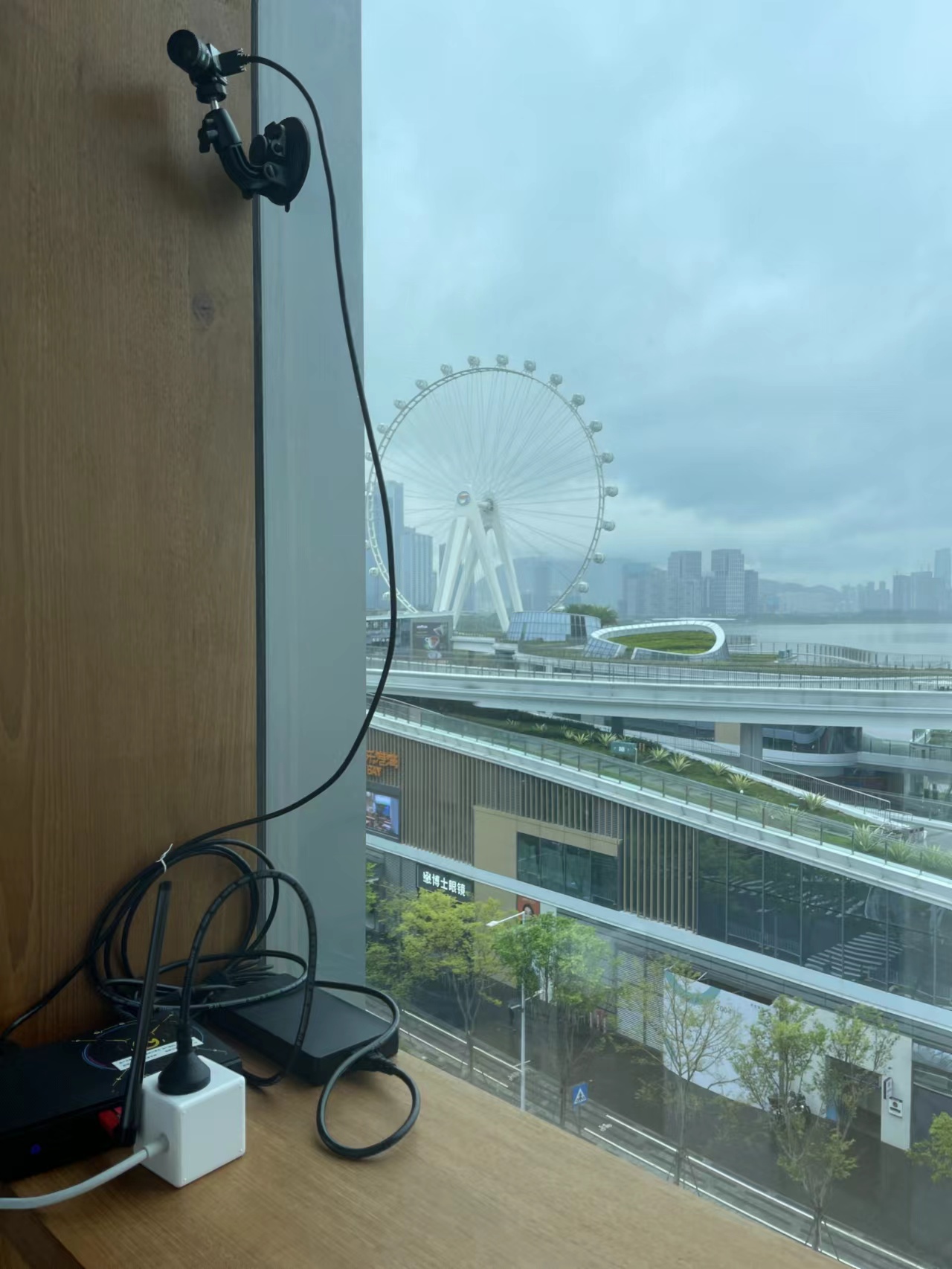}
 \caption{The data collection site.}
 \label{location}
\end{figure}

\begin{figure}[h]
 \centering
 \includegraphics[width=\columnwidth]{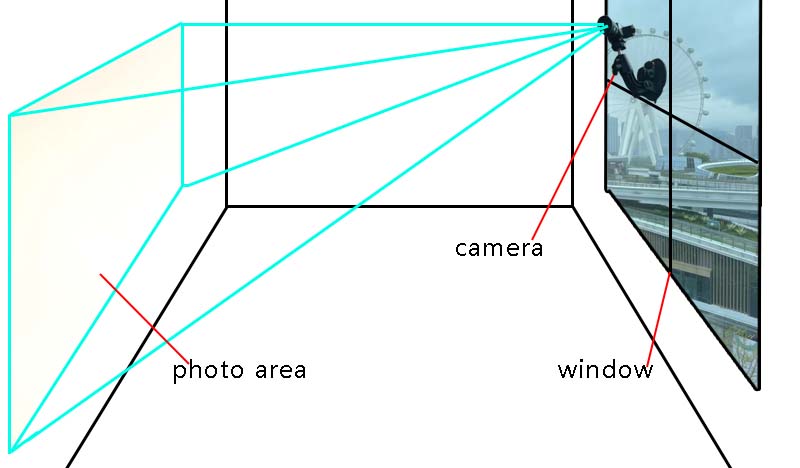}
 \caption{The sampling condition.}
 \label{condition}
\end{figure}

\begin{figure}[h]
 \centering
 \includegraphics[width=0.8\columnwidth]{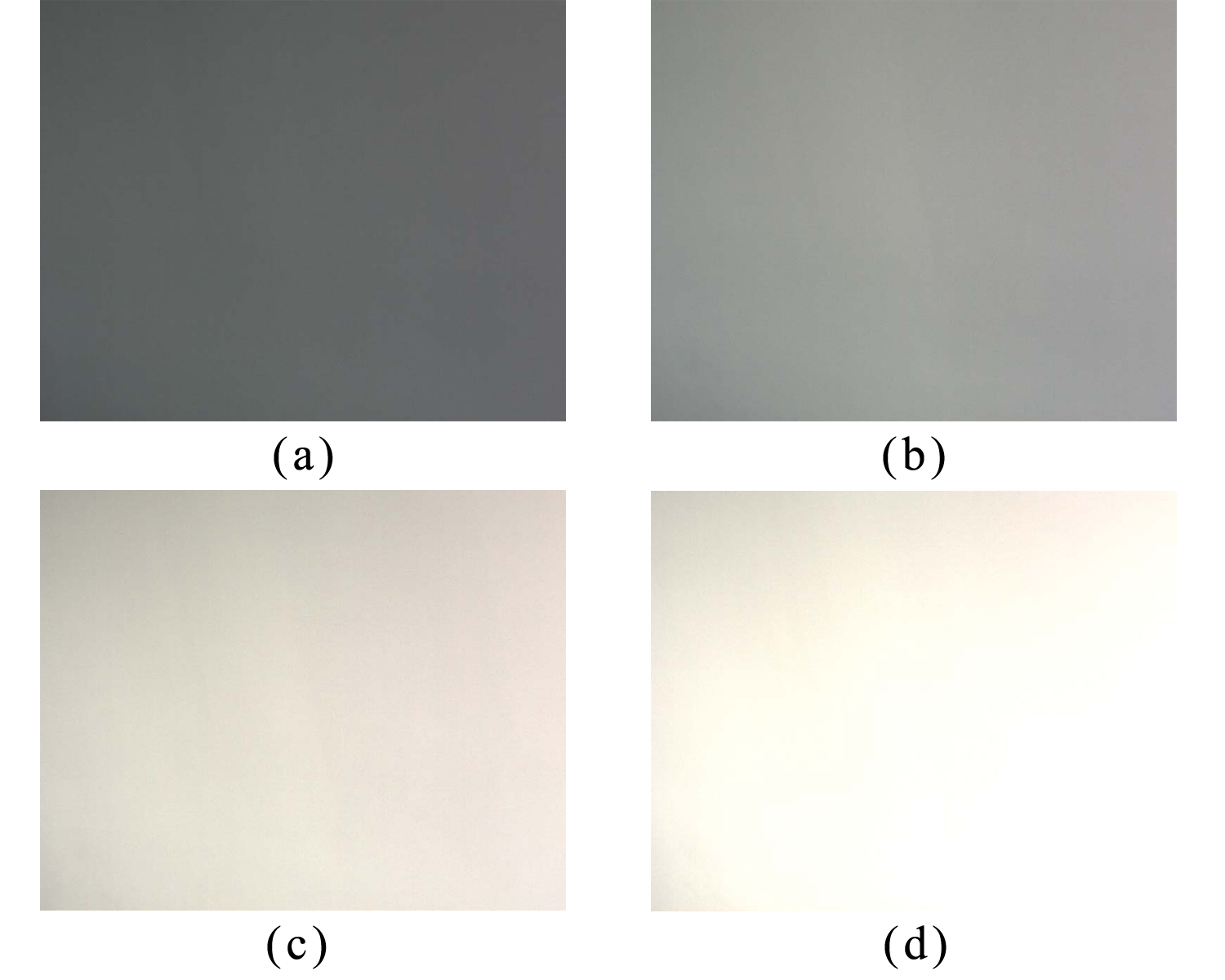}
 \caption{The sample image at (a)9:30, (b)10:30, (c)11:30, (d)12:30(black circle appears).}
 \label{data}
\end{figure}

The experiment images are taken in Ferris Gallery, a newly-founded modern art museum, located at the 6th floor of The Ocean Mansion, Baoan District in Shenzhen, 113.89°E, 22.55°N. We collect our data in one of the showrooms of the gallery, which has a white wall that faces south and a large window opposite. As shown in Fig.\ref{location}, a camera is installed on the window with a sucker located 1.5m from the wall to collect the image of the wall where the sunlight casts. We use an industrial camera with a 6 mm focal length and 1.3-million-pixel, which ensures long-term stable and reliable acquisition and high imaging quality. See Fig.\ref{condition} as a diagram of the image acquisition condition. Images are taken from 7:00 pm to 19:00 am, every 10 minutes, and are reassigned to 750×600 pixels, 300 PPI for further processing. Note that the automatic white balance function has been turned off to guarantee that the acquired images have consistent color gain. Some sample images acquired at different times are shown in Fig.\ref{data}. We collected images from April 28th to May 4th, 7 days and 546 images in total.

\begin{figure}[h]
 \centering
 \includegraphics[width=0.8\columnwidth]{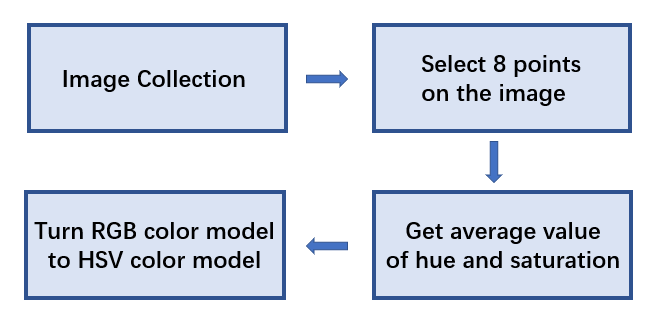}
 \caption{The pipeline of data collection.}
 \label{pipeline}
\end{figure}

\begin{figure}[h]
 \centering
 \includegraphics[width=0.8\columnwidth]{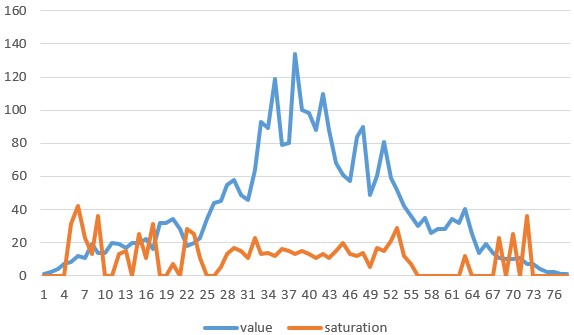}
 \caption{The change of sunlight data in one day.}
 \label{curve}
\end{figure}

Because of the wide-angle lens and the limited distance between the camera and the wall, "black circle" appears on the images, i.e., the central part of the images is brighter than the corners, see Fig. \ref{data}(d). To avoid such "black circle" and possible noises affecting extracting color information from the image, we calculate the mean color of 8 points around the centre of the image. Fig.\ref{pipeline} shows the pipeline of data collection. We calculated the average value of saturation and value in one image, and got 546 pieces of data. Fig.\ref{curve} shows the value and saturation data of May 1st.

\subsection{Data Visualization}

\subsubsection{HSV color model}

\begin{figure}[h]
 \centering
 \includegraphics[width=0.7\columnwidth]{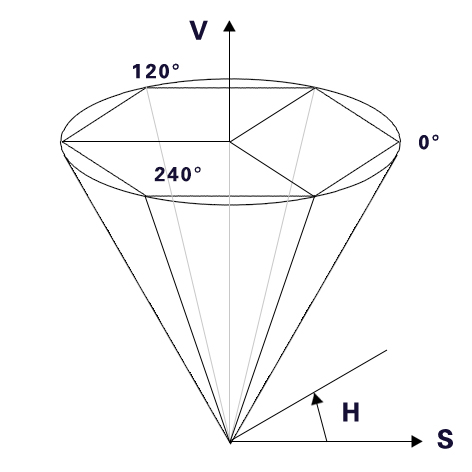}
 \caption{The HSV color model.}
 \label{hsv}
\end{figure}

Usually, digital color images are represented with the RGB color model, with each pixel described by the intensity of the red (R), green (G) and blue (B) components. Compared with commonly used RGB color model, HSV color model is closer to how humans perceive color. It has three components: hue, saturation, and value, describing colors (hue or tint) in terms of their shade (saturation or amount of grey) and their brightness value. HSV color model also mimics the way an artist mixes paints on their palette. An artist will choose a pure hue, or pigment, and lightens it to a tint of that hue by adding white or darkens it to a shade of that hue by adding black \cite{Smith:1978:CGT}. HSV color model can be illustrated by an inverted hexcone model, as shown in Fig. \ref{hsv}.  In the hexcone model, hue corresponds to color circle, defined as an angle in the range $[0, 2\pi]$ where red starts at 0, green starts at $2\pi/3$, and blue starts at $4[\pi]/3$. Saturation corresponds to the purity of the color, decrease saturation will increase whiteness. In the hexcone model, it is measured as a radial distance from the central axis to the outer surface, which starts at 0 and ends at 1. If the saturation is 0, the color becomes grey, if saturation is 1, the color becomes the purest at the given hue and value. Value is also called intensity, which is analogous to shining a white light on a colored object. The range of value is from 0 to 1, where 0 is completely black, and 1 is the brightest and reveals the most color. In the hexcone model, at the central axis, the intensity decreases from top to down, and the color changes from the darkest to the lightest \cite{Sural:2022:SHG}.

Compared to the color cube representation of RGB color model, HSV color model is more intuitively satisfying and convenient to an artist. In HSV color model, each color can be represented as a point on the hexcone. So, the visualization of color data in HSV color model can be regarded as a deformation of the hexcone model. For the above reasons, we decided to use HSV color model for representation of the sunlight data in our proposed visualization method. 

Transformation from RGB color model to HSV color model can be achieved as follows \cite{Smith:1978:CGT}.

Given: $R, G, B$, each in the domain $[0, 1]$

Desired: The equivalent $H, S, V$, each on range $[0, 1]$

\begin{enumerate}
    \item $V = \max(R, G, B)$;
    \item Let $X = \min(R, G, B)$;
    \item $S = \frac{{V - X}}{V}$; if $S = 0$ return;
    \item Let $r = \frac{{V - R}}{{V - X}}$, $g = \frac{{V - G}}{{V - X}}$, $b = \frac{{V - B}}{{V - X}}$;
    \item If $R = V$ then $H = ($if  $G = X$  then  $5 + b$ else  $1 - g)$;
    \item If $G = V$ then $H = ($if  $B = X$  then  $1 + r$ else  $3 - b)$;

    \item Else $H = ($if $ R = X$ then $3 + g$ else  $5 - r)$;
    \item $H = \frac{H}{6}$.
\end{enumerate}

\subsubsection{Visualization model}

\begin{figure}[h]
 \centering
 \includegraphics[width=0.7\columnwidth]{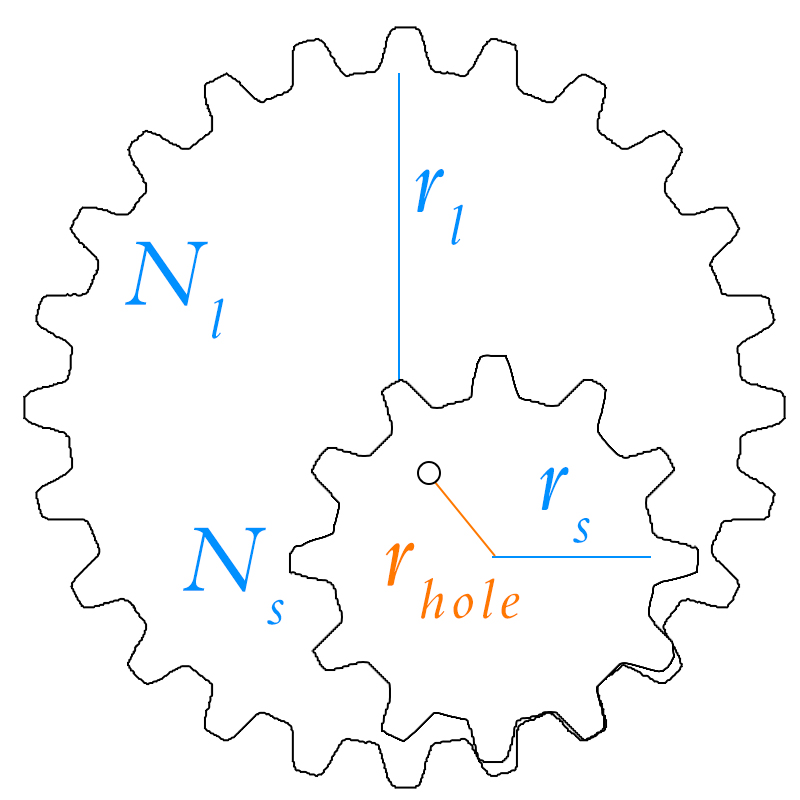}
 \caption{The gear model.}
 \label{gear}
\end{figure}

We proposed a model for the data visualization of the sunlight color data in HSV color model, which maps the saturation and value of HSV color model to the position of hole and radius of the small gear in the gear model. The visualization model is composed of a larger gear and a small gear. The number of teeth in the larger gear is fixed and greater than the number in the smaller gear. The number of teeth in the smaller gear is defined as the first variable,$N_s$, of the model. While the smaller gear rotates and revolves inside the larger gear, the sizes of the teeth in both gears are fixed and keep identical. During this procedure, the teeth of the smaller gear always cling to the teeth of the larger one. There is a hole in the smaller gear. The second variable, $r_hole$, is defined as the radial distance from the central axis to the hole, as illustrated in  Fig.\ref{gear}.

While the small gear rotates, a dynamic curve is created via the hole. Assume that the number of teeth in the larger gear is $N_l$, the period of the pattern is calculated as $[N_l, N_s]$. Assume that the radius of the larger gear and the small gear are $r_l$ and $r_s$, respectively. The number of teeth in the smaller gear is calculated as:
\begin{equation}
    N_s = \lceil N_l \cdot \frac{r_s}{r_l} \rceil
\end{equation}

Since $N_s$ needs to be an integer, the result is rounded up. From the equation, if $N_l$ is fixed, the ratio and period only depend on the radius of the small gear. We expect the period to be large to create a more complicated pattern. To make $[N_l, N_s]$ larger, $N_l$ needs to be a prime number. Meanwhile, the smallest distance between the edge of the larger gear and the pattern is determined by $r_{hole}$. So, under this assumption, the shape of the pattern depends on $N_s$ and $r_{hole}$.

We put the inverted hexcone to an inverted cone to let every vertex of hexcone be tangent to the cone. So, the HSV color model can be analogous to an inverted cone. In the following part, the inverted cone model will be used for visualization. Each point on the inverted cone can represent a color in HSV color model.

\paragraph{Value}
In HSV color model, colors with different values will be mapped to different points on the inverted cone, where the height is different. Each cross-section that is horizontal to the base of the cone is also circular. Since the correspondence between the radius of the cross-section and the height is unique, the radius of cross-section for colors with different values will be different. We assign the radius of the base to $r_l$, and the radius of the cross-section to $r_s$ because the radius of any cross-section is smaller than the radius of the base in the cone. In this way, the value of the base color of the HSV color model is mapped to $r_l$, and the value of any chosen color is mapped to $r_s$  of the proposed model. Because the period of the pattern is uniquely determined by $r_s$, colors with different values will create different patterns, and we can calculate $r_s$. Assume that the value of the base is 1, the value at the cross-section is $v$, then $r_s$ is:

\begin{equation}
    r_s = r_l \cdot v
\end{equation}

where v is the value of the color. And we can further calculate $N_s$ as:

\begin{equation}
    N_s = \lceil N_l \cdot v \rceil
\end{equation}

In this way, the larger the small gear is, the brighter the color is.

\paragraph{Saturation}

The radius of the small gear represents the radius of the cross-section, so the hole in the small gear can be represented as a hole in the cross-section. Every point in the inverted hexcone represents a color in HSV color model, so the small hole in the inverted hexcone can also represent a color in HSV color model, where the radial distance of the point represents the saturation of this color. In this way, colors with the same value but different saturation are mapped to the points on the same cross-section on the hexcone, where the radial distances are different. The experiment shows that the saturation value is between 0 and 0.16, to enlarge the difference, we multiply a coefficient c=2.55 since the range of color in HSV color model is 0 to 255. The radius of the smaller gear is:

\begin{equation}
    r_{hole}=r_s \cdot s \cdot c
\end{equation}

\paragraph{Hue}
The smaller gear is rotating, so the center angle of the small hole is meaningless. In the real case, the difference in hue between different color is small, so here we just eliminate the hue information.

\subsubsection{One more thing: Auralization model}
\begin{figure}[h]
 \centering
 \includegraphics[width=\columnwidth]{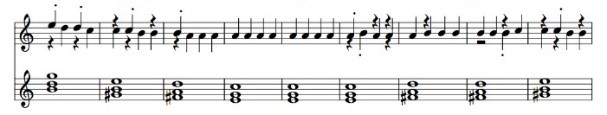}
 \caption{The staff.}
 \label{staff}
\end{figure}

The change of color data mainly varies in value; the changes in saturation and hue are relatively small. In the auralization process, we mainly use value data. We convert the saturation data and map them to the c3-f5 region as the main melody. This region creates the most pleasant melody due to the mean opinion score of 15 volunteers. Specifically, we first map the value data into the range of [0, 255], and then take every 15 values as one pitch, and map them from c3 to f5 to construct the main melody. Since the sampling interval is identical, the time value of each tone is converted to 1 second. In order to enrich the arrangement, we set 4 beats to a bar, and the major triad is added to the first note of each bar, as shown in Fig.\ref{staff}.

\subsubsection{Results}
\begin{figure}[h]
 \centering
 \includegraphics[width=\columnwidth]{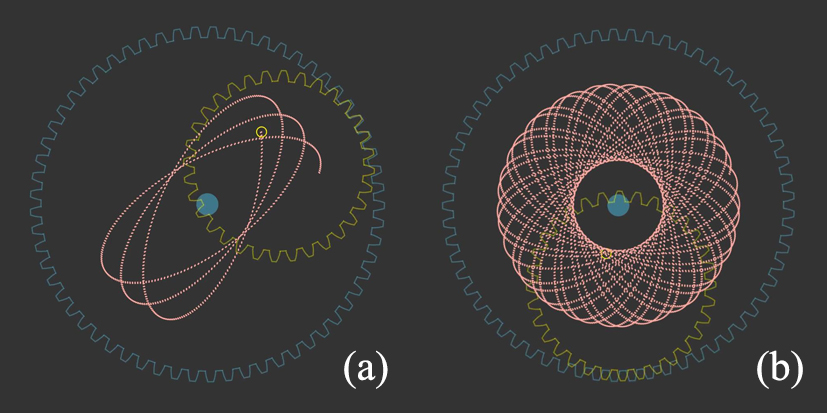}
 \caption{Result of color s=0.12, v=0. 5. (a) The pattern after 30 second, (b)The pattern after 90 seconds.}
 \label{r}
\end{figure}

\begin{figure}[h]
 \centering
 \includegraphics[width=\columnwidth]{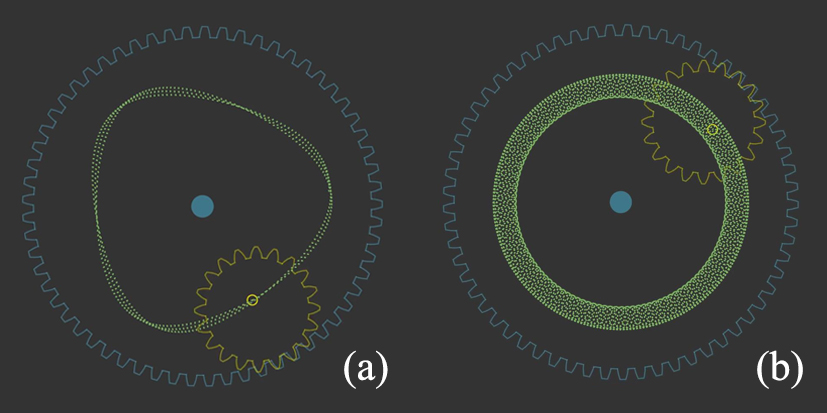}
 \caption{Result of color s=0.08, v=0.15. (a) The pattern after 30 second, (b)The pattern after 90 seconds.}
 \label{g}
\end{figure}

\begin{figure}[h]
 \centering
 \includegraphics[width=\columnwidth]{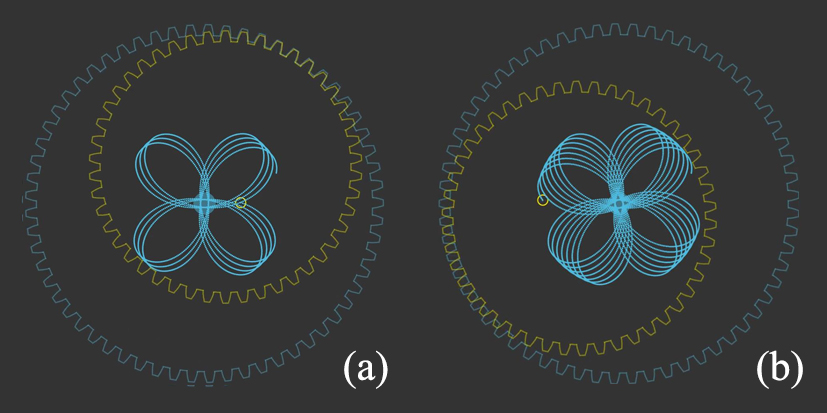}
 \caption{Result of color s=0.18, v=0.5. (a) The pattern after 30 second, (b)The pattern after 90 seconds.}
 \label{b}
\end{figure}

For each image, $N_l$ is chosen as 61. Fig.\ref{r}, \ref{g} and \ref{b} show the results of different color represented with the HSV model. From the created pattern, we can observe that when the value is larger, the smaller gear will be larger and the pattern will be concentrated around the origin of the larger one. When the saturation is larger, the radius of the pattern will have a larger coverage area. This result is like humans’ perception of color. When the intensity value is large, humans will regard the color to be bright and striking, and the pattern closer to the origin is striking at first glance. When the saturation is large, humans will regard the color to be pure, and the color of the pattern, which has a larger coverage area, looks purer than the one with a smaller coverage area.

\section{ARTWORK}
The proposed visualization model and auralization model are further applied in two public artworks named "The Mood of the Sunlight", which are two video animations using visual and audible means to express the sunlight data in an artistic way.

\subsection{Visualization artwork}
\begin{figure}[h]
 \centering
 \includegraphics[width=\columnwidth]{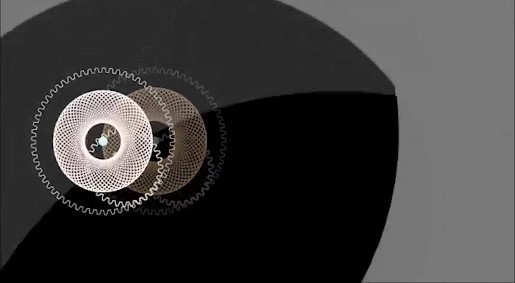}
 \caption{A frame of the visualization artwork.}
 \label{visulization}
\end{figure}

We use red, orange, yellow, green, cyan, blue and purple to draw the gears for 7 days, and each planet gear represents the change of sunlight in a day. The 7 planet gears appear and disappear in a dynamic background, and move with the dynamic background. The gears of 7 days can appear at the same time instead of appearing in sequence, showing a laterally comparison of the data in different days. Thus, emotions break through the confinement of time, showing the synchronicity of mood, and become perpetual presence. At the same time, we artificially added abnormal values to the data, causing sudden changes in the movement of the gear, breaking the law of mood, and conveying the unpredictable beauty of emotion. A frame of the visualization artwork can be seen in Fig.\ref{visulization}.

\subsection{Auralization artwork}

\begin{figure}[h]
 \centering
 \includegraphics[width=\columnwidth]{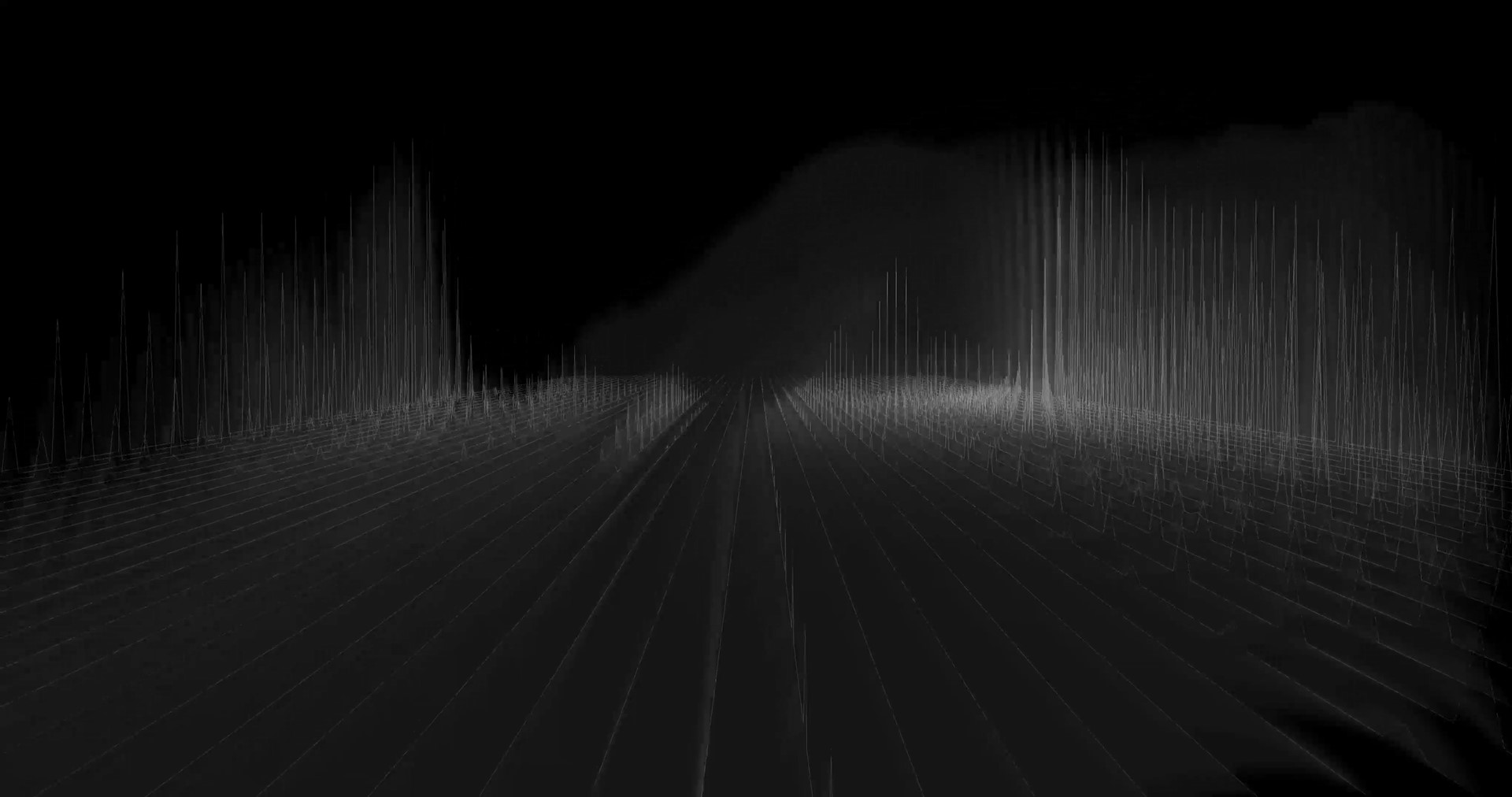}
 \caption{A frame of the auralization artwork.}
 \label{auralization}
\end{figure}

We select the data from one day to construct the music score, and use TouchDesigner to match the music score with audio images. The images are like waves, and each peak represents a pitch. When a note is played, the waves at the corresponding position will rise and fall, as shown in Fig.\ref{auralization}. Compared with the visualization artwork which conveys synchronic and unpredictable emotions, the auralization artwork emphasizes the continuity and mobility of mood. 

\subsection{Exhibition}
\begin{figure}[h]
 \centering
 \includegraphics[width=\columnwidth]{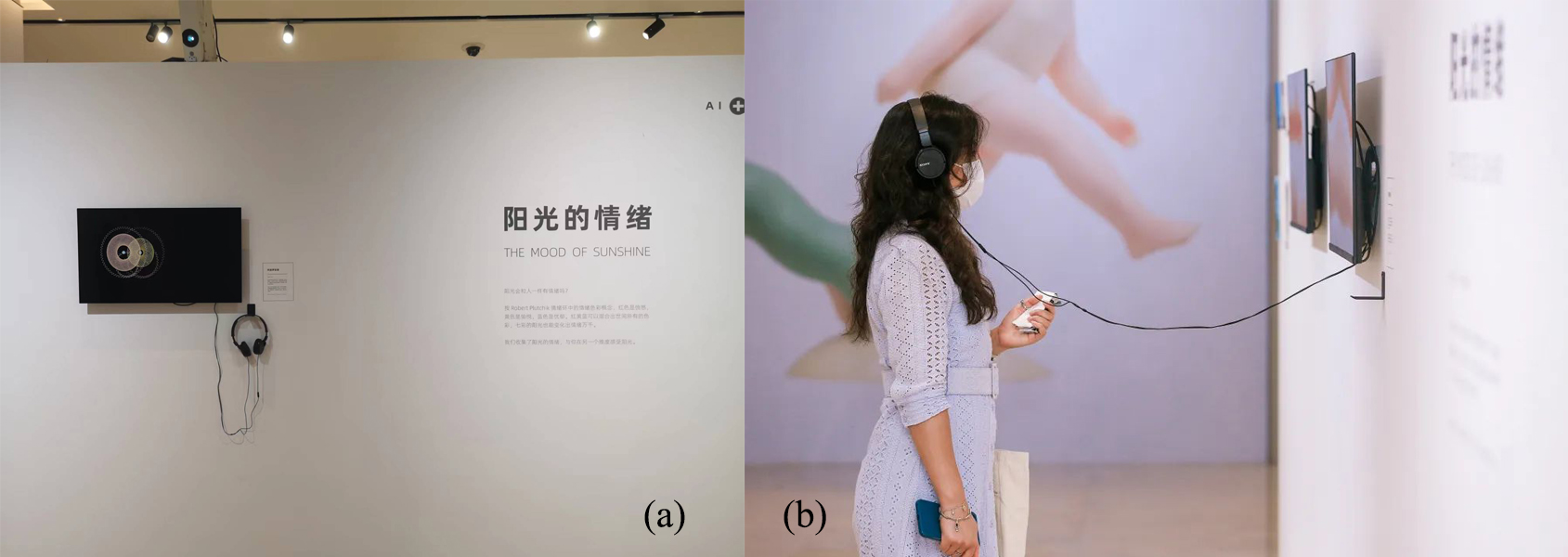}
 \caption{(a). The exhibition (b). An audience is enjoying the artwork.}
 \label{exhibition}
\end{figure}

Both the visualizaiton and auralization artworks have been on displayed at the exhibition jointly held by Shenzhen Institute of Artificial Intelligence and Robotics for Society and Shenzhen Runyi Culture Development Corporation Ltd at China Resources Tower, Shenzhen. Two LCD screens were used to play the video animations of the visualizaiton and auralization public artworks. An earphone is provided so that the audience can listen to the auralizated sunlight. Fig. \ref{exhibition} shows two pictures of the exhibition site.

\section{DISCUSSION}
Science and art may seem like two completely different disciplines, but they actually have a lot in common and can be integrated in a number of ways. Both science and art involve creativity, observation, and exploration of the world around us. Science and art can be integrated through various means, such as scientific illustration, data visualization, artistic interpretation of scientific concepts, and science-inspired art installations. Our work is a typical practice of creating public installations with data visualization technology, which provides an example of how science and art can come together.

The use of an industrial camera to collect sunlight data over a long period of time demonstrates a commitment to empirical observation and data collection. The utilization of the HSV color model to represent the variation of the "mood" of sunlight shows how the color theory can be used to represent and convey information in a visually compelling way. The creation of visual patterns with planetary motion and geometrical principle that corresponds to the changing "mood" of the sunlight reflects both scientifically accurate and artistically engaging. The use of the collected sunlight data to generate music is a way to bring the scientific data to life in a different sensory experience. 

Finally, the creation of public artworks using the visualization and auralization methods makes scientific concepts accessible and engaging to a wider audience through the power of art.

\section{CONCLUSION}

In this work, we explore the way of integrating science and public art, and develop a visualization method for the sunlight data to create a public artwork that shows the “mood” of sunlight. The visualization method generates dynamic patterns by mapping the saturation and value of the sunlight data in HSV color model to the geometric dimensions and motion of a planet gear which generates curves when rotating to present the variation of the data. By associating the variation of sunlight data with the change of human mood, we endow the sun with human emotion. Besides, we also propose an auralization model based on the change in the value of the sunlight data. Both the visualization model and auralization model are further designed as public artworks named "The Mood of the Sunlight" and has been displayed at an exhibition, allowing audiences to experience the "mood" of sunlight in a unique and engaging way. Besides, this work is hopeful to evoke deep thinking and further exploration of how science and art can come together and lead to the creation of interesting, impactful and meaningful things.
    
However, in the current form, both the visualization patterns and the auralization result are created by the data collected in advance. In the future, we will further improve the visualization and auralization methods and encapsulate them into a real-time tool so that they can reflect the real-time patterns that change with the variation in sunlight data. And we will further improve the auralization model to improve its auditory experience.

\acknowledgments{
This work is supported by the Science+Art Joint Laboratory of Benyuan Design and Research Center and Shenzhen Institute of Artificial Intelligence and Robotics for Society. Special thanks should go to Mr. Ruiqing Fu, who has put considerable time and effort into helping us to develop the software for data sampling.}

\bibliographystyle{unsrt}

\bibliography{template}

\begin{thebibliography}{10}

\bibitem{Zhang:2021:AFI}
J.~Zhang.
\newblock Analysis on the fusion of intelligent digital technology and media art.
\newblock In {\em 2nd International Conference on Intelligent Design (ICID)}, 2021.

\bibitem{Qu:2022:DVM}
M.~Qu, Y.~Sun, and Y.~Feng.
\newblock Digital media and vr art creation for metaverse.
\newblock In {\em 2022 2nd Asia Conference on Information Engineering (ACIE)}, pages 48--51, 2022.

\bibitem{Li:2022:AI}
L.~Li.
\newblock The impact of artificial intelligence painting on contemporary art from disco diffusion's painting creation experiment.
\newblock In {\em 2022 International Conference on Frontiers of Artificial Intelligence and Machine Learning (FAIML)}, pages 52--56, 2022.

\bibitem{Zou:2021:SNP}
Z.~Zou, T.~Shi, S.~Qiu, Y.~Yuan, and Z.~Shi.
\newblock Stylized neural painting.
\newblock In {\em 2021 IEEE/CVF Conference on Computer Vision and Pattern Recognition (CVPR)}, pages 15684--15693, 2021.

\bibitem{Kosara:2007:VCI}
R.~Kosara.
\newblock Visualization criticism - the missing link between information visualization and art.
\newblock In {\em 2007 11th International Conference Information Visualization (IV '07)}, pages 631--636, 2007.

\bibitem{Ceccarini:2021:EDV}
C.~Ceccarini, S.~Mirri, P.~Salomoni, and C.~Prandi.
\newblock On exploiting data visualization and iot for increasing sustainability and safety in a smart campus.
\newblock {\em Mobile Networks and Applications}, 26:2066--2075, 2021.

\bibitem{Bujari:2020:MSV}
A.~Bujari, O.~Gaggi, and C.~E. Palazzi.
\newblock A mobile sensing and visualization platform for environmental data.
\newblock {\em Pervasive and Mobile Computing}, 66, 2020.

\bibitem{OHandley:2022:CPV}
B.~J. O'Handley, M.~K. Ludwig, S.~R. Allison, M.~T. Niemier, S.~Kumar, R.~Bualuan, and C.~Wang.
\newblock Coursepathvis: Course path visualization using flexible grouping and funnel-augmented sankey diagram.
\newblock In {\em IS\&T Conference on Visualization and Data Analysis}, 2022.

\bibitem{Brewer:2022:CPM}
C.~A. Brewer, G.~W. Hatchard, and M.~A. Harrower.
\newblock Colourbrewer in print: A catalog of colour schemes for maps.
\newblock {\em Cartography and Geographic Information Science}, 30:5--32, 2022.

\bibitem{Healey:2003:CEC}
C.~Healey.
\newblock Choosing effective colours for data visualization.
\newblock In {\em Proceedings of the 7th Conference on Visualization, VIS 96}, pages 263--271, 2003.

\bibitem{Samsel:2018:AAE}
F.~Samsel, L.~Bartram, and A.~Bares.
\newblock Art, affect and colour: Creating engaging expressive scientific visualization.
\newblock In {\em IEEE VIS Arts Program (VISAP)}, pages 1--9, 2018.

\bibitem{Saito:2002:FR}
Y.~Saito.
\newblock Financial reminiscences: an example of art based on information visualization in finance.
\newblock In {\em Proceedings Sixth International Conference on Information Visualisation}, pages 759--760, 2002.

\bibitem{Brunet:2014:EDA}
K.~Brunet.
\newblock Electrotravelgram: Data art \& diy sensors.
\newblock In {\em 2014 International Conference on Virtual Systems \& Multimedia (VSMM) IEEE}, 2015.

\bibitem{Samanci:2017:FOO}
O.~Samanci and A.~Snyder.
\newblock Fiber optic ocean: Merging media for data representation.
\newblock In {\em 2017 IEEE VIS Arts Program (VISAP)}, pages 1--5, 2017.

\bibitem{Liu:2022:IDV}
X.~Liu, C.~Liu, M.~Lin, Y.~Li, J.~Xiang, and Q.~Dong.
\newblock Research on interactive design for visualization and auralization of biological data in diaphsasia.
\newblock In {\em 2022 3rd International Conference on Intelligent Design (ICID)}, pages 255--259, 2022.

\bibitem{Yu:2016:APV}
F.~Yu.
\newblock The analysis of physical visualization in artistic creation.
\newblock In {\em 2016 Nicograph International (NicoInt)}, pages 135--135, 2016.

\bibitem{Schroeder:2016:VBS}
D.~Schroeder and DF. Keefe.
\newblock Visualization-by-sketching: An artist's interface for creating multivariate time-varying data visualizations.
\newblock {\em IEEE Trans Vis Comput Graph}, pages 877--885, 2016.

\bibitem{Smith:1978:CGT}
A.~R. Smith.
\newblock Color gamut transform pairs.
\newblock {\em ACM Siggraph Computer Graphics}, 12(3):12--19, 1978.

\bibitem{Sural:2022:SHG}
S.~Sural, Gang Qian, and S.~Pramanik.
\newblock Segmentation and histogram generation using the hsv colour model for image retrieval.
\newblock {\em Proceedings International Conference on Image Processing}, 2022.

\end{thebibliography}
\end{document}